# Nuquantus: Machine learning software for the characterization and quantification of cell nuclei in complex immunofluorescent tissue images

Polina Gross[1,+], Nicolas Honnorat[2,+], Erdem Varol[2,+], Markus Wallner[1], Danielle M. Trappanese[1], Thomas E. Sharp[1], Timothy Starosta[1], Jason M. Duran[1], Sarah Koller[1], Christos Davatzikos[2*] and Steven R. Houser[1*]

[1] Temple University School of Medicine, Cardiovascular Research Center, Department of Physiology, Philadelphia, PA 19140, USA.

[2] University of Pennsylvania, Center for Biomedical Image Computing and Analytics, Department of Radiology, Philadelphia, PA, 19104, USA.

[*] Corresponding authors (Christos.Davatzikos@uphs.upenn.edu and srhouser@temple.edu)
[+] These authors contributed equally to this work

## ABSTRACT

Determination of fundamental mechanisms of disease often hinges on histopathology visualization and quantitative image analysis. Currently, the analysis of multi-channel fluorescence tissue images is primarily achieved by manual measurements of tissue cellular content and sub-cellular compartments. Since the current manual methodology for image analysis is a tedious and subjective approach, there is clearly a need for an automated analytical technique to process large-scale image datasets. Here, we introduce Nuquantus (Nuclei quantification utility software) - a novel machine learning-based analytical method, which identifies, quantifies and classifies nuclei based on cells of interest in composite fluorescent tissue images, in which cell borders are not visible. Nuquantus is an adaptive framework that learns the morphological attributes of intact tissue in the presence of anatomical variability and pathological processes. Nuquantus allowed us to robustly perform quantitative image analysis on remodeling cardiac tissue after myocardial infarction. Nuquantus reliably classifies cardiomyocyte versus non-cardiomyocyte nuclei and detects cell proliferation, as well as cell death in different cell classes. Broadly, Nuquantus provides innovative computerized methodology to analyze complex tissue images that significantly facilitates image analysis and minimizes human bias.

# INTRODUCTION

Histological evaluation of tissue samples is a prevailing diagnostic method in the study of cellular pathologies in a variety of diseases. Tissues undergoing pathological remodeling, with changes in the number and types of cells are often evaluated using immunofluorescent staining to define changes in specific cell types in complex multicellular organs. In most instances, a single tissue section is simultaneously stained for multiple cell markers. Advanced microscopy imaging[1] enables high-resolution visualization of stained tissue and acquisition of a multitude of images that require detailed analyses. However, most image analytical techniques currently used to define the number and types of cells in complex tissues, as well as the proliferative or apoptotic "state" of specific cell types, are usually subjective and time consuming.

Biomedical image processing is a progressively developing field that applies computerized approaches to facilitate and augment microscopy image analysis[2]. Well established methods have been previously applied on histopathological images to execute morphological cell analyses[3-5]. Some studies have segmented whole cells[6, 7] specifically, in isolated cell lines[8] or intact tissue sections, in which cell borders were clearly apparent[9, 10]. However, segmentation of intact cells *in situ* becomes problematic when cell boundaries are not stained and cannot be easily visualized. In addition, *in situ* cell segmentation is exceedingly challenging due to intact tissue arrangement, condensed cellular structures and the presence of heterogeneous cell populations. Further complications are observed in pathology images because the damaged tissue forms irregular cell structures and undergoes dynamic healing processes[11]. In consequence, whole cell segmentation may be an ineffective strategy for intact tissue image analysis.

Automated nuclei segmentation is another well-studied approach that has been previously applied on histopathology images[12-14]. Nuclei segmentation employs spatial information such as size, shape and texture to classify nuclei. This information can be further used to classify mitotic nuclei, abnormal chromatin distribution and irregular nuclear boundaries. While nuclei segmentation has been widely applied in cancerous cells studies[12, 15], its application in other pathology disciplines (e.g. acute tissue injury, ischemia, hypoxia) requires more intricate techniques, because after tissue injury nuclei of existing cell subtypes may morphologically change and new cell types can migrate into the damaged tissue. Some of these different cells subtypes may exhibit similar nuclei features, making global nuclei segmentation indistinguishable between their matching cell subpopulations. Therefore, general nuclei segmentation may not be adequate to quantify the information contained in injured tissue images.

Currently, there is a paucity of reliable automated tools to quantitatively analyze complex fluorescent images obtained from injured tissues. Therefore, manual subjective examination still remains the standard approach. This extremely laborious task is plagued by three major limitations: 1) *Time* – manual analysis of hundreds and even thousands of images is a meticulous task that is time consuming. Additionally, the manual workload is significantly increased if an image was acquired via multiple fluorescent channels. In this case, an image has to be dissected into multi-channels so that each channel is manually examined. 2) *Bias* – tendency of the analyst to favorably or unfavorably quantify cells or cellular components that fit one's hypothesis may severely skew results. 3) *Inconsistency* – Different analysts may quantify the information perceived from the same set of images in a way that introduces large variability and results in conflicting findings. Considering the complexity of histological tissue images and the key challenges involving the manual approach for *in situ* image examination, development of automated image analysis algorithms is a fast growing necessity.

To address the shortcomings of the computerized tissue image analysis, we have designed a new image processing software we termed Nuquantus (**Nu**clei **quant**ification **u**tility **s**oftware). Nuquantus is a novel supervised machine learning framework that segments, classifies and quantifies nuclei of cells of interest in complicated tissue images. Nuquantus is an open source Matlab program that can be downloaded from https://www.cbica.upenn.edu/sbia/Erdem.Varol/nuquantus.html.

We demonstrate the utility of Nuquantus in cardiac tissue images taken from animal models of ischemic cardiac injury, known as myocardial infarction (MI). Fluorescent histopathology images of the heart after MI illustrate the difficulty and the complexity involved in defining cardiomyocytes (CMs) and CM nuclei with computerized image processing. Firstly, common MI animal models include small and large animals in which cardiac anatomy and physiology differ. For this reason, development of generic software that is adaptive to biological differences between species is challenging. Secondly, the heart is a muscular organ in which muscle fibers are oriented in different directions across the cardiac wall, making the identification of CMs difficult. Upon heart fixation and tissue slicing into histological sections, the orientation of CMs can vary from round cross sections to transverse or longitudinal rectangular shapes. Additionally, the heart contains many different cell types, with CMs comprising less than one-third of the cells[16, 17]. In healthy cardiac tissue, these anatomical issues can complicate the identification of CMs and their corresponding nuclei. In cardiac injury, the identification of CMs and CM nuclei becomes further complicated since CMs can morphologically remodel or uncouple from their neighbors, in concurrence with massive infiltration of inflammatory cells[18].

CM nuclei can be manually classified and quantified without cell membrane specific staining. However, as stated above, this technique is time-consuming, prone to bias and can be inconsistent between different analysts. Recent studies on cardiac regeneration have greatly relied on manual measurements of the number or percentages of apoptotic CM nuclei, as well as new CM nuclei that have been recently formed[19-22]. The data from these studies can be widely variable, likely due to the subjective analytical tools that were employed.

The innovative approach of Nuquantus allows streamlining image analysis of complex tissue, such as cardiac tissue. While many other automated cell analysis methods explicitly attempt to segment cells *in situ* using information of the membrane stains, Nuquantus detects and quantifies nuclei corresponding to cells of interest even in the absence of cell membrane information. Moreover, Nuquantus allows customized image analysis of a variety of imaging datasets from different animal models or imaging protocols since it is an adaptive learning based framework. Importantly, the semi-automated pipeline of Nuquantus greatly speeds up image analysis, reduces human bias and produces reproducible data that increases consistency between different analysts.

## RESULTS

**Nuquantus design.**
The goal of Nuquantus design was to identify and classify specific cell subtype nuclei in tissue sections with cellular diversity. In this study, we portray the application of Nuquantus in fluorescent stained cardiac tissue images.

Nuquantus is a sequential pipeline of image processing modules that takes raw histological images and outputs the desired quantification of nuclei of cells of interest. The initial step in the Nuquantus pipeline (Fig. 1) includes image preprocessing to correct microscopy illumination artifacts (Fig. 1b and Supplemental Fig. S1 online). All nuclei present in the image are segmented and then approximately hundreds of high-level shape and texture descriptors are extracted from all nuclei and their surroundings (Fig. 1c). Next, prediction of nuclei of cells of interest within the heterogeneous pool of cell populations is made based on patterns learned a priori from a training (labeled) set of images via supervised learning models (Fig. 1d). To further refine the predictions, the user can confirm the automated predictions and rule out possible false positives (Fig. 1e). Lastly, a final output with classified cell specific nuclei is provided to the user along with a quantitative summary (Fig. 1f). The quantitative analysis includes colocalization information of the selected nuclei with additional molecular markers. For instance, additional markers may include specific stainings for DNA synthesis, DNA fragmentation, cytosolic or nuclear proteins.

We exemplify Nuquantus classification performance of CM nuclei in composite cardiac tissue images after ischemic myocardial injury. We used histological samples from two different MI animal models: a large animal model for ischemia reperfusion in swine, and small animal model with permanent coronary artery occlusion in mice (see methods). Histological observation of the cardiac tissue reveals morphological variations between the two animal models of post MI remodeling (Fig. 2). These variations are attributable to a variety of factors including differences in heart size and MI experimental protocols[23, 24]. We identified normal viable CMs as nucleated cells positive for α-sarcomeric actin – a cardiac specific structural protein. Nuclei that were not found in the center of α-sarcomeric actin staining represent nuclei of interstitial cells that reside within the cardiac tissue. These nuclei belong to fibroblasts, endothelial cells, vascular smooth muscle cells and immune cells[17, 23]. Three sub-regions of the heart were compared: I) Infarct Area (IA) as the site of ischemic injury where most CMs and supportive tissue die and are eventually replaced by scar (Fig. 2a); II) Border zone (BZ) as the interface between the infarct area and the viable tissue with mixtures of CMs and scar (Fig. 2b), and III) Viable zone (VZ) (Fig. 2c). The IA and BZ contain a mixture of cell types along with variation in CM size and abnormal shape. These differences exist between the three cardiac sub-regions within the same animal model. Moreover, the features of the same cardiac sub-region differ between the large and small animal models, making the overall CMs identification very difficult.

To achieve accurate CM nuclei segmentation in images with high anatomical and morphological variability, we trained a specific machine learning model for each type of animal (Fig. 3). An expert cardiac histologist manually labeled 84 images of mouse post-MI cardiac tissue (26 images of BZ, 34 images of VZ and 24 images of IA). In these images, 16,957 nuclei were labeled out of which 1,673 were identified as CM nuclei. Also, 88 images of swine post-MI cardiac tissue (35 images of BZ, 37 images of VZ and 16 images of IA) were labeled. These images included 19,162 nuclei labels out of which 3,361 were labeled as CM nuclei. All the images were pre-processed as detailed in the method section. Common cell segmentation features such as Haralick[25] were extracted for each nucleus. In addition, novel features such as Fourier histograms of gradients (HOG)[26] and cell membrane detector (CMD)[27, 28] were implemented for the first time, to our best knowledge, for the purpose of nucleus segmentation. These extracted features quantitatively captured the shape and the texture information of the nucleus and its surrounding environment to mimic the features recognized by the histologist when attempting to detect CM nuclei. Extracted features of all nuclei – CM and non-CM nuclei were input into a logistic regression model[29] that learned the patterns of the ground truth provided by the expert cardiac histologist's labels. Next, predictions on unseen nuclei (leave-image-out training as described in methods) were computed via probability scores in order to discriminate a non-CM nucleus from a CM nucleus (Supplemental Fig. S2 online). Based on these scores, we generated a precision-recall curve to assess the performance of each animal prediction model.

For both the mouse and swine models, the logistic regression thresholds were determined by leave-one-image-out cross-validation. For the swine model, we selected a logistic regression threshold corresponding to CM nuclei detection at 90% recall and 75% precision (Fig. 3a). With this approach, for every 20 true CM positive nuclei, 5 nuclei may be classified as a false positive. On the other hand, for every 20 true CM positive nuclei, only 2 CM nuclei will be lost as false negatives. For mouse CM nuclei detection, we selected a threshold of 85% recall and 32% precision (Fig. 3b). The interpretation of this precision-recall trade-off is that for every true positive detected CM nucleus, 2 false positive nuclei will be detected. The false positive nuclei can then be removed in the optional user correction step. Image analysis using Nuquantus contests the manual approach as a result of substantial image clean-up. Given that mouse images contain up to 90% non-CM nuclei, as provided by the histologist's ground truth, Nuquantus automatically removes 80% of non-CM nuclei. The remaining clean-up to be carried out by the user correction step is estimated to be only 20%.

The difference in precision between the mouse and swine models is explained by the relatively more complicated image textures found in mouse images. Mouse cardiac tissue histological specimens contain CMs at a variety of planes of section (cross, longitudinal section and oblique section) within the same slide, due to the small overall heart size. Histological specimens from large animal models are more likely to have CMs at only one plane of section, making analysis more uniform. Nevertheless, the selected high recall for the mouse model successfully reduces the loss of false negative CM nuclei. Furthermore, we tested the model performance on different cardiac sub-regions: VZ, BZ and IA to observe the region specific performance of the prediction models. The prediction of CM nuclei in the mouse BZ performs at higher precision-recall (52%, 85% respectively) than VZ or IA. While IA is the region where the majority of CMs die due to the injury, the BZ remains a relatively enriched zone with CMs. Nevertheless, the CM density in BZ is less compact in comparison to the VZ. Hence, segmentation of individual CM nuclei in the BZ becomes an easier task than in IA or VZ.

Swine CMs nuclei detection is shown in Fig. 3c. The model reliably identifies the vast majority of the true positive CM nuclei, while filtering out most of the non-CM nuclei.

**Nuquantus is a rapid and reliable image analysis method.**

The effectiveness and the efficiency of CM nuclei segmentation by Nuquantus were compared to the standard approach of manual CM nuclei identification (see methods). We computed inter-observer agreement between four independent cardiac histologists that manually labeled CM nuclei in a set of five VZ of mouse cardiac tissue images (Fig. 4a). Then, same four histologists used Nuquantus to segment CM nuclei in a new set of five VZ of mouse cardiac tissue images. A different set of images was used in each experiment to avoid bias towards previously seen images. However, the complexity of the cardiac tissue in all images was preserved. The inter-observer agreement between the histologists in each experiment was computed using the Dice overlap measure (see methods). The mean agreement between the histologists using the manual approach was 64 ± 15% (mean ± standard deviation), the median agreement was 63% and the coefficient of variation was 23%. The mean agreement between the histologists using Nuquantus was 70 ± 5% (mean ± standard deviation), the median agreement was 72% and the coefficient of variation was reduced to 8%. With a confidence level of 90%, we observed an improvement of the inter-observer agreement using Nuquantus. Subsequently, the CM nuclei segmentation via Nuquantus software was more consistent across different users with significantly reduced variance when compared to the standard manual approach (Fig. 4a).

Another important improvement introduced by Nuquantus was the analysis speed up of CM nuclei segmentation in complex tissue images. To demonstrate the rate of image analysis, a total of 36 mouse cardiac VZ images were analyzed by four histologists and the time of analysis was recorded for each image. Half of the images were analyzed by Nuquantus and the other half was manually analyzed. CM nuclei segmentation using Nuquantus significantly expedited the average time for image analysis. On average, one image analysis via Nuquantus was completed in 2.2 ± 0.9 minutes (mean ± standard deviation) versus 12.6 ± 5.6 minutes via manual analysis (Fig. 4b).

In summary, Nuquantus accelerated the image analysis by 6 fold and reduced the variability of the CM nuclei segmentations produced by the different histologists.

**Cell proliferation and apoptosis quantified by Nuquantus**

Cell proliferation with synthesis of new DNA takes place in tissue post injury. Cell proliferation in the post-MI heart occurs primarily in non-CM interstitial cells and potentially in CMs[17, 30, 31]. Detection of proliferating cells is possible via 5-ethynyl-2′-deoxyuridine (EdU) – a thymidine analog, which is incorporated into the DNA of a cell nucleus during DNA replication phase. The incorporated EdU reacts with azide–modified fluorescent dye that can be detected by confocal imaging[32, 33]. In our study design, we implemented EdU minipumps in a mouse *in vivo* MI model (see methods). Detection of DNA replicating cell nuclei in the heart was observed 1 week post-MI in IA, BZ and VZ (Fig. 5a). Nuclei positive for DAPI staining (blue) and EdU staining (green) surrounded by positive α-sarcomeric actin (red) defined CM nuclei undergoing DNA synthesis. Nuclei positive for DAPI and EdU staining but not surrounded by positive α-sarcomeric actin were defined as non-CM nuclei undergoing DNA synthesis. Overall, nuclei subjected to DNA replication were largely detected in IA and BZ. However, the existence of clumped nuclei and lack of clear CM cell borders overburden the quantification and classification of EdU and DAPI positive CM nuclei versus non-CMs nuclei. Additionally, the inhomogeneous light illumination, a known limitation of confocal imaging, overwhelms the identification of any EdU and DAPI positive nuclei in the dark edges of the image.

To quantitatively evaluate Nuquantus performance given the above mentioned challenges, we compared between manual classified nuclei counts obtained by a histologist, and Nuquantus counts obtained after image intensity correction and nuclei segmentation following the user correction step (Supplementary Fig. S3 and Fig. S4 online). The ratio of Nuquantus segmented CM nuclei out of total nuclei was similar in all post-MI and sham zones when compared to manual approach (Fig. 5b). As expected, Nuquantus identified a gradually decreasing percentage of CM nuclei in BZ and IA compared to VZ, suggesting that the site of injury contained a reduced number of CM cells. An opposite trend was observed with the percentage of nuclei undergoing DNA synthesis. Similarly to manual approach, Nuquantus confirmed that the percentage of proliferating cells with positive EdU and DAPI nuclei were primarily found in IA (56% manual vs. 44% software) and BZ (50% vs. 42%), and secondarily in VZ (26% vs. 27%) (Fig. 5c). In comparison to the manual counts, Nuquantus shows slightly increased detection of proliferating CM nuclei (Fig. 5d and 5e), which can be attributed to the image preprocessing and illumination correction. These findings are consistent with the pathogenesis of injured tissue, in which the site of injury becomes an attractive platform for proliferating inflammatory cells, active angiogenesis, scar tissue formation and potentially tissue regeneration. Notably, almost no proliferation was measured in sham mice (< 3% using manual and Nuquantus approaches), confirming that these animals did not suffer from any ischemic injury in the heart.

A controversial question in the cardiac field is whether adult CMs are capable of proliferating after MI. Out of cell subtypes that survived the infarct injury in the IA and the newly infiltarating cell subtypes to the region of the insult, only ~1% CM nuclei were identified (Fig. 5b). From this ~1% of CM nuclei, 16% were identified by manual counts and 31% were identified by Nuquantus as containing positive nuclei for EdU and DAPI staining (Fig. 5d), indicating DNA synthesis. Out of ~5% CMs found in the BZ, only 4% vs. 8% (manual vs. Nuquantus, respectively) contained nuclei undergoing DNA synthesis. Among ~12% of CMs found in the VZ, only 2% vs. 6% (manual vs. Nuquantus, respectively) CM nuclei were undergoing DNA synthesis. In shams, both manual and Nuquantus methods indicated < 1% EdU+DAPI positive CM nuclei. No statistical difference was obtained between the reported proportions using both approaches (2-sample z-test). In a comprehensive scale, Nuquantus confirmed that less than 1% of nuclei positive for EdU+DAPI staining were classified as CM nuclei in IA (Fig. 5e). A similar tendency was observed in BZ and VZ (< 2%) via manual and Nuquantus methods.

These findings suggest that Nuquantus reliably detects nuclei undergoing DNA synthesis and is capable of classifying these nuclei into CM versus non-CM nuclei. Quantification of these data by Nuquantus matches the standard manual approach and confirms that DNA synthesis in CM nuclei is a low frequency event in the post-MI heart.

CMs undergo programmed cell death (noted apoptosis) after an ischemic injury[34, 35]. Histological detection of apoptosis is possible with Fluorometric TUNEL assay (see methods) that identifies nuclear DNA fragmentation [36]. To validate Nuquantus for detection and classification of nuclei undergoing DNA fragmentation, we obtained tissue images from healthy mouse hearts. In those hearts, no apoptosis or very low physiological rate of apoptosis should be evident. Therefore, these images represent TUNEL negative controls. In positive TUNEL control (see methods) extensive DNA fragmentation is detected indicating a vast apoptosis (Fig. 6a). Staining for sarcomeric tropomyosin was performed to detect CMs. We were able to validate Nuquantus classification of CM nuclei using our mouse training model (Fig. 6B and Supplementary Fig. S5 online). TUNEL positive CM nuclei were determined if nuclei were positive for DAPI staining (blue) and TUNEL staining (green) surrounded by positive sarcomeric tropomyosin staining (red).

In contrast to TUNEL negative controls where the percentage of total apoptotic nuclei remained very low: 0.09% vs. 0.04% (manual vs. Nuquantus, respectively), in the positive TUNEL control the percentage of total nuclei was elevated to 96% vs. 88% (manual vs. Nuquantus, respectively) (Fig. 6c). Similar results were obtained when comparing the percentage of apoptotic CM nuclei in negative and positive TUNEL controls – manually versus Nuquantus segmentation. No statistical difference in percentages obtained between manual and Nuquantus approaches was found (2-tailed 2-sample z-test).

These studies show that, Nuquantus successfully classified CM and non-CM nuclei undergoing DNA fragmentation. Quantification output of Nuquantus was validated in comparison to the quantification using the manual approach.

## DISCUSSION

Gaining histological insight into cellular mechanisms that are driving diseases requires efficient analysis tools to reliably detect and quantify sub-cellular components of interest. Yet, a major bottleneck exists in the analysis of complex fluorescence tissue images, which is often dependent on subjective visual assessment and manual labeling. Despite advances made with automated cell and nuclei segmentation[6, 12], the practical challenge remains to segment cells in intact tissue without visual cell boundaries, and to identify specific cell of interest nuclei in the presence of diverse cell subtypes.

In this study, we have trained Nuquantus, a supervised machine learning software, to classify specific cell subtype nuclei, namely CM nuclei in fluorescent tissue images of remodeling heart. Nuquantus training was based on CMs that exhibited mature sarcomeric structure. In overall, Nuquantus extracts hundreds of high-level image features from which it learns patterns of tissue architecture. Subsequent to learning informative patterns in these features, Nuquantus automatically distinguishes between nuclei subclasses – CM versus non-CM nuclei via computation of probability scores. The advantage of this approach is that CM nuclei are classified in an objective computerized fashion rather than relying on subjective human perception. Previously, it was shown that combining information across microscopy multi-fluorescence channels substantially improved the identification of the cellular components in complex tissue images[37]. Therefore, Nuquantus integrates information from the segmented nuclei (blue channel) and their cellular surroundings (red channel) to achieve accurate nuclei classification. To avoid machine learning overtraining[38], we used different subsets of images to train Nuquantus. Images from IA, BZ and VZ were applied during Nuquantus training to make the system robust to variability present in the post MI heart images within the same animal model. This way, we ensured that the training model can generalize and adapt to new incoming images without being systematically biased towards specific type of images. In the present study, we have trained CM nuclei classification in two MI animal models: one for small animals (mice) and one of large animals (swine). Ideally, a unified prediction model would be preferable to identify CM nuclei in all animal models. However, the scale and CM morphology differences are too large to learn consistent patterns that can generalize in all animal types.

Post MI cardiac remodeling elucidates typical structural and cellular modifications that occur in tissue pathology: cellular rearrangement, fibrosis, cell death, cell regeneration, inflammation, angiogenesis, etc. Often these cellular modifications affect multiple cell populations, which in this case are CM and non-CMs[17]. For this reason, specific CM nuclei identification is highly challenging. Although, fully accurate CM classification may not be attainable, we have optimized the classification precision and recall in each animal model, such that an optimal trade off is maintained between truly classified CM nuclei and potentially misclassified CM nuclei. An ideal classifier would have an absolute separation between CM and non-CM predicted nuclei. However, nuclei that reside in convoluted tissue regions, such as IA and BZ, may be classified as false positives or false negatives. The opted high recall essentially eliminates true non-CM nuclei from the predictions and significantly facilitates image analysis. In addition, the corresponding level of precision enables detection of the vast majority of true CM nuclei, and can be further boosted by the user in the semi-automated correction step. Although somewhat labor consumig, semi-automated user correction step focuses on reviewing nuclei segmentation results rather than manually annotating images. Indeed, Nuquantus has been shown to significantly speed up image analysis, decrease the analysis workload and reduce variability between multiple observers.

Cardiac regeneration through new CM formation remains an area of controversy[39-41], partially due to shortage of objective quantitative tools. In this work, Nuquantus was validated against standard approach of manual image analysis. We conclude that Nuquantus is a consistent strategy for image analysis in a quantitative reproducible manner. Notably, measurements of CM proliferation and CM apoptosis ratios are computed after user interaction step, making the introduction of human bias minimal. The user interactive step focuses only on verification of

CM nuclei classification while masking out any additional fluorescent channels containing information regarding DNA synthesis (EdU stain) or DNA fragmentation (TUNEL stain). Eventually, the quantification of EdU or TUNEL positive CM nuclei is objective and fully automated. Nevertheless, automated quantification of EdU positive CM nuclei does not determine the source of the putative new CMs and does not distinguish between CM proliferation versus CMs deriving from a precursor (stem) cell.

The limitation of Nuquantus in the present study is that the image analysis is tailored to the same type of images that the supervised training was performed on. As a consequence, the analysis of images obtained on either different magnification or different tissue subtypes requires training of another machine learning model. Morphological diversity exists between different types of tissue and organs. Also, image acquisition at either higher or lower magnification alters the amount and the quality of feature information that can be gathered from a single image. However, the modular design of Nuquantus and its ultimate feature toolbox allow the users to train a new machine learning model to accommodate to their needs and specifications of their imaging datasets (also available at the software website).

In the future, Nuquantus will be applied to other tissue immunohistochemistry protocols such as Hematoxylin and eosin stain (H&E stain) and immunofluorescence combined with cell membrane stains. Additional immunohistochemistry techniques are required to identify small immature CMs. While these staining protocols are in progress, Nuquantus can be further trained to classify subgroup of newly forming CMs. In addition, Nuquantus will be upgraded to allow interactive machine learning model[42]. In this type of online machine learning, the user's interactive input, is incorporated into the existing prediction model in real time to further improve the predication accuracy.

In summary, Nuquantus is a free, open source, Matlab software that is customizable for user needs and could be used for annotating cell nuclei in a large variety of histology images, up to the most challenging ones. The machine learning approach of Nuquantus allows cell subtype specific nuclei classification in complex fluorescence tissue images in an accurate, objective and rapid manner. Nuquantus is a high-throughput software, which enables semi-automated image analysis, and increases the amount of the data that can be quickly and accurately processed.

# METHODS

**Implementation.** The Nuquantus software was developed in Matlab 2014a and is freely available for research purposes. The access to Nuquantus software is provided in the following website: https://www.cbica.upenn.edu/sbia/Erdem.Varol/nuquantus.html. The experiments presented in this paper were carried out on Windows 7 - 64 bit, 4GB RAM, Intel i5 2.5GHz CPU.

**Animals and myocardial infarction**. All animal procedures were approved and performed in accordance with the guidelines and regulations of the Temple University School of Medicine Institutional Animal Care and Use Committee. Permanent occlusion myocardial infarction (MI) was induced in 12-week old male C57BL/6 mice (The Jackson Laboratory; Bar Harbor, ME) by ligation of the left anterior descending coronary artery (LAD) as previously described[43, 44]. Sham operated animals underwent surgery without LAD ligation. After recovering from surgery, osmotic mini pumps (Azlet; Cupertino, CA) containing 39 mg/mL 5'-ethynyl-2'-deoxyuridine (EdU) (Life Technologies; Carlsbad, CA) were implanted subcutaneously between the two scapulae to deliver a continuous infusion of EdU (Life Technologies; Carlsbad, CA) at a flow rate of 0.5 µL/hr for 1 week. Mini pumps containing only vehicle (50/50 DMSO/ddH2O) were also implanted in control mice group. All pumps were removed after 1 week; mice were sacrificed and the hearts were extracted via cardioectomy. Gottingen female mini-swine age of 7-9 months (Marshall BioResources, NY, USA) were used for angioplasty catheterization to temporarily occlude the blood flow in the LAD and induce ischemia reperfusion MI model[45]. The swine were sacrificed 1 month post-MI and cardioectomy was performed. All invasive procedures were performed under anesthesia as previously described[44, 45].

**Tissue Processing.** In mice, the heart was arrested in diastole by intracoronary perfusion of 100 mM cadmium chloride and 1 M potassium chloride. Both mouse and pig hearts were gravity perfused with 10% formalin, processed and embedded into paraffin wax blocks. Wax blocks were cut into 5 µm thick sections and mounted onto glass slides (AML Laboratories, Baltimore, MD, USA).

**Immunofluorescence histochemistry.** Slides underwent deparaffinization for antigen retrieval with citrate buffer. The following primary antibodies were used: α-sarcomeric actin (A2172, Sigma-Aldrich; St. Louis, MO), sarcomeric tropomyosin (Sigma-Aldrich #T9283; St Louis, MO). Secondary antibodies were used to detect the primary antibodies as followed: rhodamine red-X donkey anti-mouse IgM (Jackson Immunoresearch Laboratories; West Grove, PA) to detect α-sarcomeric actin or sarcomeric tropomyosin. Nuclei were stained with 4', 6-diamidino-2-phenylindole (DAPI, Millipore; Billerica, MA). EdU detection was performed using the EdU Click-iT Imaging kit (Life Technologies; Carlsbad, CA).
Immunostaining with terminal deoxynucleotidyl transferase-mediated deoxyuridine triphosphate nick-end labeling (TUNEL) was performed on paraffin-embedded heart tissues obtained from12 weeks old wild-type male C57BL/6 mice. These mice were subcutaneously injected with 200µl of physiological solution saline and sacrificed 24 hours later. No heart injury is expected to be obtained after saline injections. The DeadEnd Fluorometric TUNEL system was used to label apoptotic nuclei (Promega #G3250; Madison, WI). For positive control slides were treated with DNASE I (10U/l) prior to TUNEL staining.

**Image acquisition.** Confocal images were randomly acquired in the infarct area, border zone, and viable (remote) zone from 19 mice and 4 swine MI animal models. 5-15 images were acquired using at least 3 slides per 1 animal. Each slide contained all 3 post-MI cardiac zones. Nikon Eclipse T1 confocal microscope (Nikon Inc; Mellvile, NY) at 40X magnification was used for image acquisition.

**Image Processing.** The images were corrected for lightning inhomogeneity and variation of tissue depth inside the field of view. These artifacts were removed by dividing, at each pixel, the intensity measured in the different channels by the local background intensity, that was measured by applying a Gaussian filter of scale 10 pixels to the red (α-actin/tropomyosin staining) channel[46]. After this intensity correction, the nuclei were segmented by extracting the connected components[47] on the blue DAPI channel. Nuclei smaller than 5 pixels were filtered out as noise.

*Feature extraction* - Five sets of features were extracted for each blue (DAPI staining) and red (α-actin/tropomyosin staining) pixel: (i) rotationally invariant histograms of gradients (Fourier HOG) [26], (ii) Haralick features[25], (iii) local intensity entropy, (iv) the output of a cell membrane detector (CMD) based on regularized steerable filters[27] and tensor voting[28] and (v) the size of the extracellular component containing the pixel (ECS). Further details are provided in the supplementary material. Then, four descriptive statistics: mean, median, standard deviation and median absolute deviation were computed to summarize feature information for each nucleus. These features were scaled independently to zero mean and unit variance for each image, in order to produce a set of measures that can be compared across images.

*Nuclei Classification* - Expert cardiac histologist tagged the nuclei that are inside cardiomyocytes to serve as ground truth labels. Total of 16,957 mouse nuclei and 19,162 swine nuclei were labeled. The labeled mouse and swine nuclei were used for training a logistic regression model using the Liblinear[29] package to obtain a mouse and a swine nuclei prediction model, respectively. Classification results were thoroughly cross-validated: the performance of the model was assessed by sequentially leaving out the nuclei from one image, termed unseen image, and training a logistic regression model on the remaining images (leave-image-out procedure). The model was applied to the nuclei in the left out image, in order to estimate their likelihood to be CM nuclei. This procedure was repeated for every single image in the training dataset to result in prediction probability scores for all nuclei. The predications of the different models were compared to the labels provided by the expert histologist for generating precision/recall curves and quantifying precisely the prediction performance of the models. The precision and recall were computed using the following formulas:

$$Precision = \frac{True\ Positive\ nuclei}{True\ positive\ nuclei + False\ positive\ nuclei}$$

$$Recall = \frac{True\ positive\ nuclei}{True\ positive\ nuclei + False\ negative\ nuclei}$$

Once an appropriate operating point for the model was chosen based on the precision/recall curves, the trained prediction models were then used to identify CM nuclei in new images whose features were extracted using the same above steps.

**Manual counts of EdU, TUNEL, DAPI, and Myocytes.** Cells from all images were manually counted in Adobe Photoshop CS6 Extended (Adobe Systems Inc.; San Jose, CA). For each image, DAPI positive staining was labeled with a dot label and counted for the total number of DAPI positive cells. To be considered EdU or TUNEL positive, cells must have contained both EdU/TUNEL and DAPI signals. In a separate photoshop layer, myocyte nuclei were labeled and counted if the nuclei were positive for DAPI and surrounded by α-sarcomeric actin or tropomyosin positive staining. All DAPI negative cells were disregarded. EdU or TUNEL positive myocytes were counted if the cell contained all three stains (EdU/TUNEL, α-sarcomeric actin/tropomyosin, and DAPI).

**Dice inter-observer agreement.**
The agreement between the expert histologists was measured by (1) matching their provided manual labels, and (2) by matching their segmented nuclei using Nuquantus software with user correction step. Manual labels and segmented nuclei were matched when the distance between their geometric centers was smaller than 10 pixels, the radius of the largest nucleus in our database. The agreement was measured as the proportion of labels or segments that match between two expert histologists (A and B) [48]: $\frac{|A \subset B| + |B \subset A|}{|A| + |B|}$, where |A| denotes the total number of labels provided by histologist A, and |A⊂B| are the labels provided by histologist A that can be matched with the labels provided by histologist B. The agreement was computed for the six possible combinations of four expert histologists. The mean agreement was reported ± standard deviation.

**Statistical analysis**

Non-parametric Wilcoxon test was used for mean comparisons. Brown – Forsythe test was used for equality of variances. For comparison of sample proportions, 2-sample z-test was applied. P value is indicated where applicable.

## FIGURE LEGENDS

**Figure 1. Nuquantus pipeline using example of a mouse cardiac tissue image**. **a.** Original immunofluorescent image of mouse cardiac tissue obtained by confocal microscopy (scale 20µm). **b.** Image preprocessing to correct and normalize illumination artifacts. **c.** Extraction of image features including texture, shapes, local disorganization, size of the extracellular space, and cell edges. **d.** Nuclei selected by Nuquantus based on predictions made by a priori trained machine learning model. If some false positive nuclei were rendered in this step, they can be discarded by the user in the following optional step. **e.** Optional user interaction step in which the user approves identified nuclei of cells of interest (cardiomyocytes) – circled in yellow. **f.** Final output with classified nuclei of cells of interest (cardiomyocytes). **g.** Quantitative output is provided with counted nuclei of different cell sub-groups.

**Figure 2. Immunofluorescent staining of cardiac tissue post myocardial infarction (MI).** Top panels: mouse MI cardiac left ventricular tissue, bottom panels: swine MI cardiac left ventricular tissue. **a.** Viable zone (VZ) (also known as remote zone) represents the tissue that is distant from the ischemic injury and contains morphologically normal CMs. **b.** Border zone (BZ) represents the transition from injury site to surviving CMs. The surviving CMs in the swine BZ are localized in adjacent form of islands whereas in mouse the CMs are equally distributed along the BZ. **C.** Infarct Area (IA) represents the site of ischemic injury that is characterized by CMs loss and formation of scar tissue. Due to differences in cardiac anatomy and MI surgical procedures, the swine IA is very large and contains interstitial cell nuclei without any CMs. On the contrary, the mouse contains few surviving CMs, which are outnumbered by the interstitial cell nuclei. Blue: nuclei staining via 4', 6-diamidino-2-phenylindole (DAPI). Red: cardiac specific structural protein staining for α-sarcomeric actin. Examples for CM nuclei are shown via white arrows and are identified if they contain DAPI positive staining and surrounded by α-sarcomeric actin positive staining. Examples for non-CMs DAPI positive nuclei are shown via yellow arrows. Scale 25µm.

**Figure 3. Performance of the swine and mouse prediction models. a.** Swine model obtained precision recall curve that describes the trade-off between precision and recall. This model was trained on 88 cardiac images of post MI tissue obtained from N=4 swine. **b.** Mouse model obtained precision recall curve. The training was performed on 84 cardiac images of post MI tissue obtained from N=19 mice. The contribution of VZ, BZ and IA images to the prediction model is shown by dashed lines. The overall performance of all cardiac sub regions is shown in red continuous line. **c.** Example of swine prediction model on unseen images. The prediction was compared to the ground truth labels to identify true positive versus true negatives, false positive and false negative CM nuclei.

**Figure 4. a. Inter-rater agreement.** The agreement of CM nuclei segmentation by different histologists was compared between the manual standard approach and the Nuquantus software after user correction step.
# P value = 0.10 using paired, one-tailed, non-parametric Wilcoxon test for mean agreements. ¥ P value <0.05 using Brown – Forsythe test for equality of variances. **b. Time analysis for CM nuclei segmentation.** Histologist's average time for CM nuclei segmentation per image was compared between manual approach and

Nuquantus software. **\*\*\*** P value < 0.001 using unpaired, one-tailed, non-parametric Wilcoxon test for mean time analysis.

**Figure 5. Nuquantus detection and quantification of nuclei undergoing DNA synthesis. a.** Cell nuclei positive for EdU (green) are extensively detected in IA and BZ and moderately in VZ. The abundance of non-CM EdU positive nuclei in these images makes the identification of CM EdU positive nuclei difficult. Scale 20µm. **b-e.** Quantitative analysis of CMs and cell proliferation was compared between the manual approach and Nuquantus software in post MI mouse model. CM nuclei were quantified only if these nuclei contained DAPI labeling. The percentage of EdU + CM nuclei was quantified as the portion of CMs nuclei co-labeled with DAPI and EdU out of all CM nuclei labeled with DAPI or all nuclei co-labeled with DAPI and EdU. Total of 48 images were analyzed from MI mice (N=4) and sham mice (N=4). No statistical difference was obtained between the reported proportions using both approaches (2-tailed 2-sample z-test).

**Figure 6. Nuquantus detection and quantification for nuclei undergoing DNA fragmentation. a.** Cell nucleus positive for TUNEL (green) is a rare event in healthy normal cardiac tissue – TUNEL negative control (upper image). Extensive TUNEL positive nuclei are observed in the TUNEL positive control (lower image) after a healthy cardiac tissue section was pre-treated with DNASE I. TUNEL positive nuclei are observed in CM and non-CM nuclei. Scale 20µm. **b.** Validation of Nuquantus CM nuclei classification in cardiac tissue stained for a different CM structural protein than the structural protein Nuquantus was originally trained for. **c.** Comparative quantification analysis of Nuquantus versus standard manual approach. CM nuclei were quantified only if these nuclei contained DAPI labeling. In total, 30 images obtained from N=3 mice were analyzed. No statistical difference was obtained between the reported proportions using both approaches (2-tailed 2-sample z-test).

**ACKNOWLEDGMENTS**

This study was supported by grants from NIH/NHLBI, HL108806 and NIH/NHLBI, HL091799.

**AUTHOR CONTRIBUTIONS STATEMENT**

PG, NH and EV equally contributed to this work. PG, NH and EV designed the software and planned the experiments. PG performed the experiments, analyzed the data, prepared the figures and wrote the main manuscript. NH designed the software features, contributed to the design of statistical analysis and wrote parts of the manuscript methods. EV wrote the software toolbox, contributed to the design of statistical analysis and wrote parts of the manuscript methods. MW provided the TUNEL mouse histological samples and participated in the experiments performance. DMT provided the EdU mouse histological samples and participated in the experiments performance. TES provided the swine histological samples and carried out the myocardial infarction surgical procedures in swine. TS contributed to the image acquisition and participated in the experiments performance. JD provided mouse histological samples. SK carried out myocardial infarction surgical procedures in mice. CD and SRH supervised the work. All authors reviewed the manuscript.

**ADDITIONAL INFORMATION**

**Competing financial interests:** The authors declare no competing financial interests.

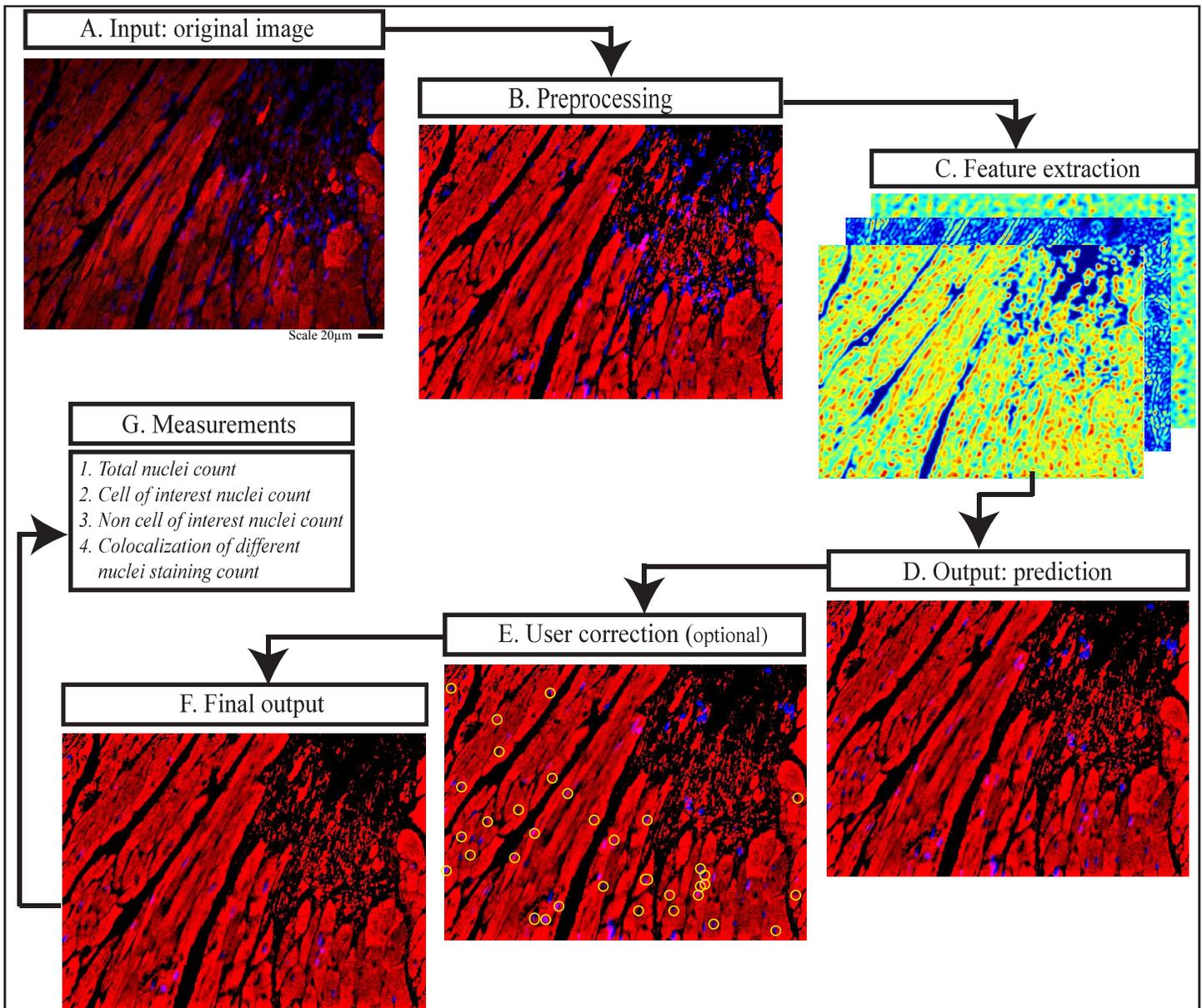

Figure 1

Figure 1. Nuquantus pipeline using example of a mouse cardiac tissue image. A. Original immunofluorescent image of mouse cardiac tissue obtained by confocal microscopy at 40X magnification (scale 20μm). B. Image preprocessing to correct and normalize illumination artifacts. C. Extraction of image features including texture, shapes, local disorganization, size of the extracellular space, and cell edges. D. Nuclei selected by Nuquantus based on predictions made by a priori trained machine learning model. If some false positive nuclei were rendered in this step, they can be discarded by the user in the following optional step. E. Optional user interaction step in which the user approves identified nuclei of cells of interest (cardiomyocytes) – circled in yellow. F. Final output with classified nuclei of cells of interest (cardiomyocytes).  G. Quantitative output is provided with counted nuclei of different cell sub-groups.

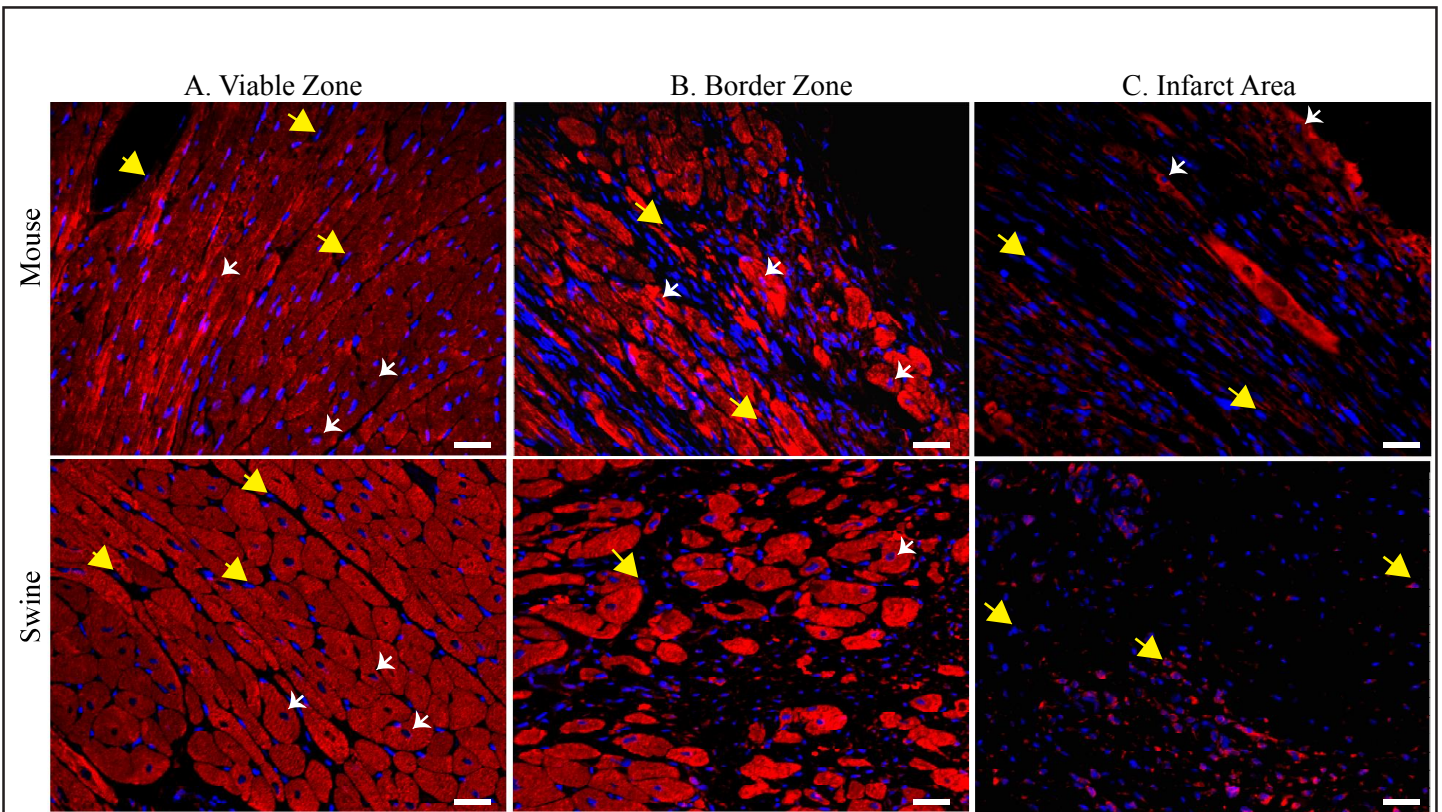

Figure 2

Figure 2. Immunofluorescent staining of cardiac tissue post myocardial infarction (MI). Top panels: mouse MI cardiac left ventricular tissue, bottom panels: swine MI cardiac left ventricular tissue. A. Viable zone (VZ) (also known as remote zone) represents the tissue that is distant from the ischemic injury and contains morphologically normal CMs. B. Border zone (BZ) represents the transition from injury site to surviving CMs. The surviving CMs in the swine BZ are localized in adjacent form of islands whereas in mouse the CMs are equally distributed along the BZ. C. Infarct Area (IA) represents the site of ischemic injury that is characterized by CMs loss and formation of scar tissue. Due to differences in cardiac anatomy and MI surgical procedures, the swine IA is very large and contains interstitial cell nuclei without any CMs. On the contrary, the mouse contains few surviving CMs, which are outnumbered by the interstitial cell nuclei. Blue: nuclei staining via 4', 6-diamidino-2-phenylindole (DAPI). Red: cardiac specific structural protein staining for α-sarcomeric actin. Examples for CM nuclei are shown via white arrows and are identified if they contain DAPI positive staining and surrounded by α-sarcomeric actin positive staining. Examples for non-CMs DAPI positive nuclei are shown via yellow arrows. Scale 25μm, 40X magnification.

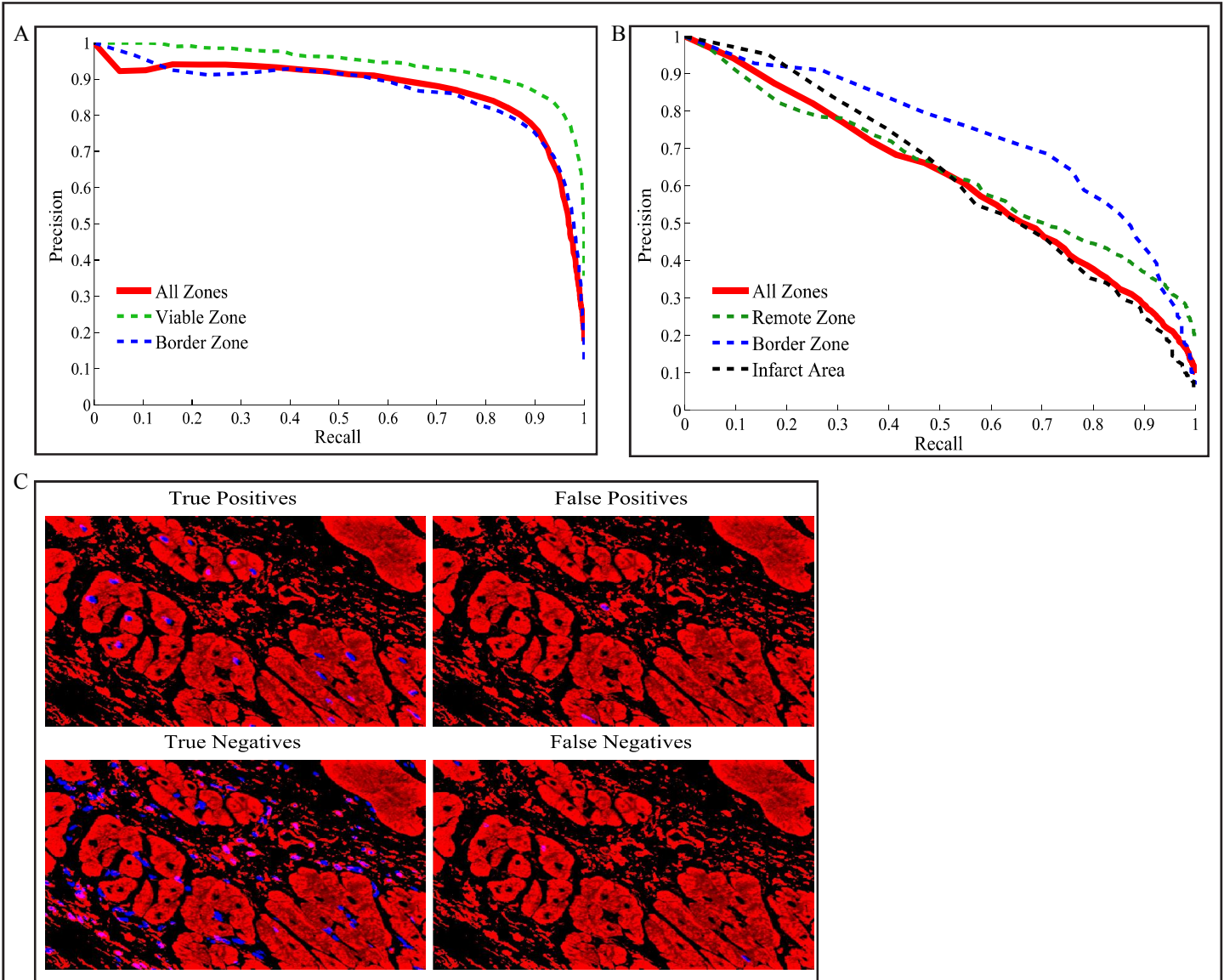

Figure 3

Figure 3. Performance of the swine and mouse prediction models. A. Swine model obtained precision recall curve that describes the trade-off between precision and recall. This model was trained on 88 cardiac images of post MI tissue obtained from N=4 swine. B. Mouse model obtained precision recall curve. The training was performed on 84 cardiac images of post MI tissue obtained from N=19 mice. The contribution of VZ, BZ and IA images to the prediction model is shown by dashed lines. The overall performance of all cardiac sub regions is shown in red continuous line. C. Example of swine prediction model on unseen images. The prediction was compared to the ground truth labels to identify true positive versus true negatives, false positive and false negative CM nuclei.

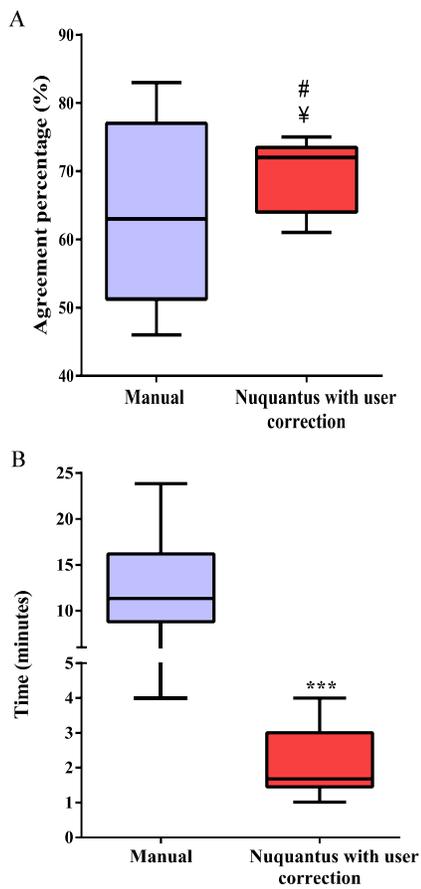

Figure 4

Figure 4. A. Inter-rater agreement. The agreement of CM nuclei segmentation by different histologists was compared between the manual standard approach and the Nuquantus software after user correction step.
# P value = 0.10 using paired, one-tailed, non-parametric Wilcoxon test for mean agreements. ¥ P value <0.05 using Brown – Forsythe test for equality of variances. B. Time analysis for CM nuclei segmentation. Histologist's average time for CM nuclei segmentation per image was compared between manual approach and Nuquantus software. *** P value < 0.001 using unpaired, one-tailed, non-parametric Wilcoxon test for mean time analysis.

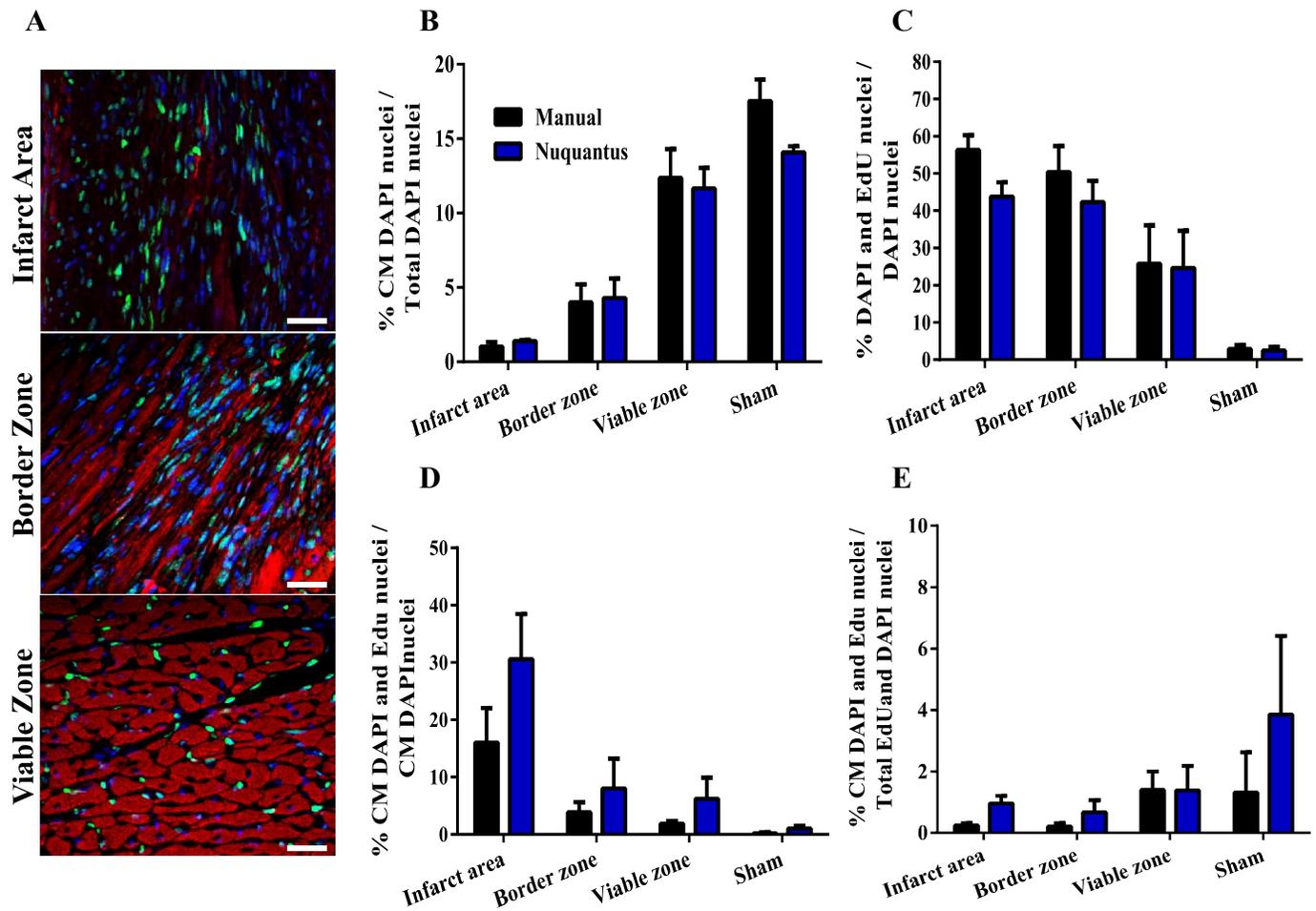

Figure 5

Figure 5. Nuquantus detection and quantification of nuclei undergoing DNA synthesis. A. Cell nuclei positive for EdU (green) are extensively detected in IA and BZ and moderately in VZ. The abundance of non-CM EdU positive nuclei in these images makes the identification of CM EdU positive nuclei difficult. Scale 20μm, 40X magnification B-E. Quantitative analysis of CMs and cell proliferation was compared between the manual approach and Nuquantus software in post MI mouse model. Total of 48 images were analyzed from MI mice (N=4) and sham mice (N=4). No statistical difference was obtained between the reported proportions using both approaches (2-tailed 2-sample z-test).

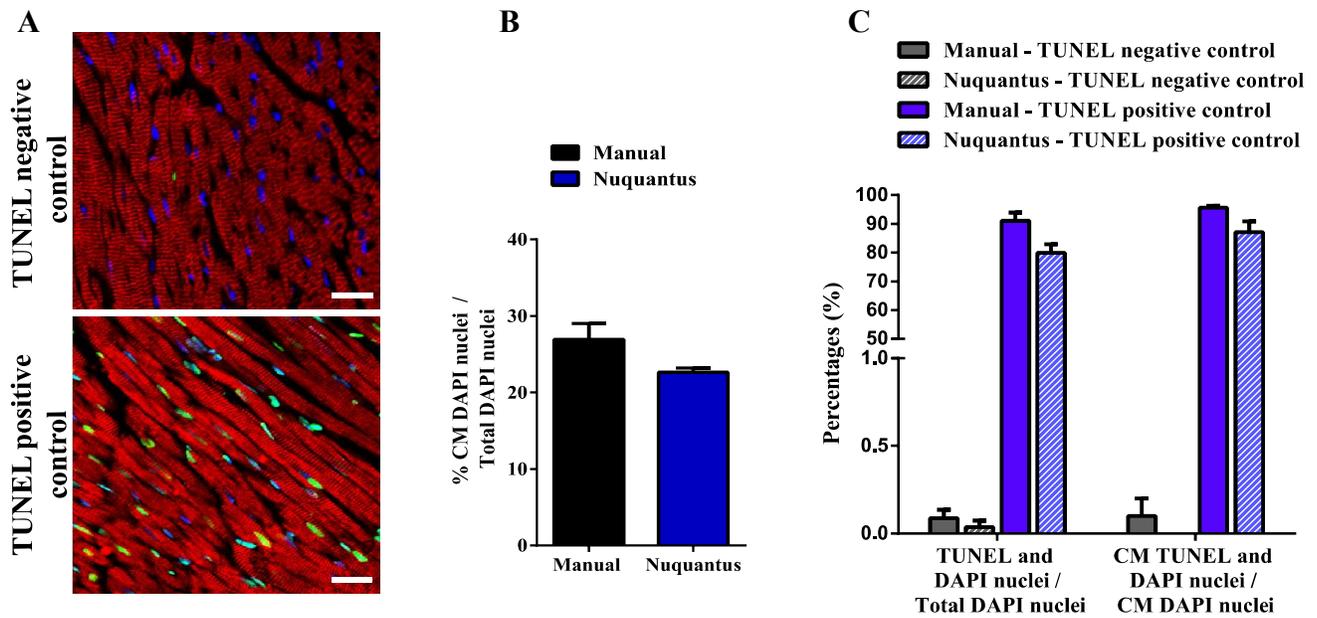

Figure 6

Figure 6. Nuquantus detection and quantification for nuclei undergoing DNA fragmentation. A. Cell nucleus positive for TUNEL (green) is a rare event in healthy normal cardiac tissue – TUNEL negative control (upper image). Extensive TUNEL positive nuclei are observed in the TUNEL positive control (lower image) after a healthy cardiac tissue section was pre-treated with DNASE I. TUNEL positive nuclei are observed in CM and non-CM nuclei. Scale 20µm, 40X magnification. B. Validation of Nuquantus CM nuclei classification in cardiac tissue stained for a different CM structural protein than the structural protein Nuquantus was originally trained for. C. Comparative quantification analysis of Nuquantus versus standard manual approach. In total, 30 images obtained from N=3 mice were analyzed. No statistical difference was obtained between the reported proportions using both approaches (2-tailed 2-sample z-test).